\documentclass[preprint,authoryear,5p,times,twocolumn]{elsarticle}

\usepackage{txfonts}
\usepackage{graphicx}
\usepackage{longtable}
\usepackage{color}
\usepackage{multirow}
\usepackage{natbib}
\usepackage{pdfpages}
\usepackage{subfigure}

\usepackage{savesym}
\savesymbol{iint}
\savesymbol{iiint}
\savesymbol{iiiint}
\savesymbol{idotsint}
\usepackage{amsmath}
\restoresymbol{AMS}{iint}
\restoresymbol{AMS}{iiint}
\restoresymbol{AMS}{iiiint}
\restoresymbol{AMS}{idotsint}
\bibpunct{(}{)}{;}{a}{}{,}
\newcommand{\beq}{\begin{equation}}
\newcommand{\eeq}{\end{equation}}

\newcommand{\bdi}{\begin{displaymath}}
\newcommand{\edi}{\end{displaymath}}

\usepackage{lineno,hyperref}
\modulolinenumbers[5]

\journal{Astroparticle Physics}

\begin{document}

\begin{frontmatter}

\title{Analysis methods to search for transient events in ground-based Very High Energy $\gamma$-ray astronomy}

%% Group authors per affiliation:
\author[lab1]{F.~Brun}
\ead{francois.brun@cea.fr}
\author[lab2]{Q.~Piel}
\author[lab3]{M.~de Naurois}
\author[lab4]{S.~Bernhard}

\address[lab1]{IRFU, CEA, Universit\'{e} Paris-Saclay, F-91191 Gif-sur-Yvette, France}
\address[lab2]{Laboratoire d'Annecy de Physique des Particules, Universit\'{e} de Savoie, CNRS/IN2P3, F-74941 Annecy, France}
\address[lab3]{Laboratoire Leprince-Ringuet, Ecole Polytechnique, CNRS/IN2P3, F-91128 Palaiseau, France}
\address[lab4]{Institut f\"ur Astro- und Teilchenphysik, Leopold-Franzens-Universit\"at Innsbruck, A-6020 Innsbruck, Austria}

\begin{abstract}
  Transient and variable phenomena in astrophysical sources are of
  particular importance to understand the underlying gamma-ray emission processes.  In the
  very-high energy gamma-ray domain, transient and variable sources are related
  to charged particle acceleration processes that could for instance help
  understanding the origin of cosmic-rays. The imaging atmospheric Cherenkov
  technique used for gamma-ray astronomy above $\sim 100$ GeV is well suited for
  detecting such events. However, the standard analysis methods are not optimal
  for such a goal and more sensitive methods are specifically developed in this
  publication.
  The sensitivity improvement could therefore be helpful
  to detect brief and faint transient sources such as Gamma-Ray Bursts.
\end{abstract}


\end{frontmatter}

%\linenumbers

%________________________________________________________________

\section{Introduction}\label{intro}

Transient or variable astrophysical sources are particularly interesting to
understand charged particle acceleration processes in the Universe. They are of
great importance to study the origin of cosmic-rays or to test the
validity of some of the most fundamental laws of physics such as Lorentz
invariance. The most violent events in the Universe such as Gamma-Ray Bursts
(GRBs) or outbursts from Active Galactic Nuclei (AGNs) accelerate particles to
high energies which are expected to radiate in High and Very-High Energies (HE:
$> 100$~MeV, VHE: $> 100$~GeV) gamma-rays.

At VHE, the Imaging Atmospheric Cherenkov Telescopes (IACTs) detect and
study variability at minute to week time scales from a variety of astrophysical
sources such as AGN, pulsars or binary systems \citep[e.g.][focusing on the AGN
  PKS 2155-304]{pks2155}. GRBs are intensely studied since their discovery in
1967 at all wavelengths~\citep[e.g.][]{Swiftobs,Fermiobs,Fermiobsgbm}. However,
due to the limited number of expected photons and the limited field-of-view and
duty cycles of IACTs, detection of such objects at VHE is challenging.
VHE gamma-ray emission from GRBs has recently been detected for the first time
by H.E.S.S. \citep{HESSGRB_2019} and MAGIC \citep{MAGICGRB_2019}. Increasing the
detection statistics of GRBs in the VHE gamma-ray domain is now crucial to
understand the processes at play in these objects.

The determination of the statistical significance of a VHE gamma-ray source in
IACT analysis is usually done by accumulating data towards a test position and
comparing the number of detected events at this position to the number obtained
in (source-free) background control
regions~\citep[e.g.][]{lima,bergebg}. However, in case of a transient source,
this method is not optimal since the test position may only contain signal
events during a limited fraction of observation and only background events
afterwards. Some methods, specifically developped to search for transient
events, are explored in this article. These methods aim at improving the
sensitivity of IACT data analysis to such events and at recovering the detection
of some of the weak and brief signals that standard analyses could miss.

The methods presented in this document have been implemented in
the analysis frameworks used to
analyze data from the H.E.S.S. experiment \citep{Combined}.  They will also be
useful for the data analysis of the next major instrument : the Cherenkov
Telescope Array \citep[e.g. ][]{CTA}.

The statistical tests we describe rely on a good knowledge of the instrument
response at each time and observed position and its evolution with time. Therefore, in section
\ref{section:acceptance}, we first describe the methods used (i) to estimate the
detector acceptance and (ii) to construct a new time reference in which the event rate is expected to be constant.
Once this step is done,
a set of (acceptance corrected) times is available, to which statistical tests
can be applied to search for variability. The tests we developed or adapted from the litterature to search
for transient events are described in section \ref{section:tests}
and their performances are estimated and compared in section
\ref{section:fakeinjection}.

%__________________________________________________________________

\section{Acceptance estimation and correction} \label{section:acceptance}

The tests presented here are based on the knowledge of the events arrival times
and the detector response. The individual event times are stored during the
analysis together with their nature (gamma-like or hadron-like events) according
to the reconstruction algorithm.
Gamma-like events are events that are reconstructed by the analysis software as
potential gamma rays due to their image properties.
In regions were no sources are present, these events
are mainly originating from hadrons or electrons. Hadron-like events on the
other hand are the events that are identified by the reconstruction method as
 the least similar to gamma-ray events.
For a steady source or for background events, the arrival times are supposed to
follow a Poisson process with constant rate. However, this behavior is affected by the detector
response variations with time. For instance, if the acceptance of the instrument
increases (due to a lower zenith angle or better atmospheric conditions), the
event rate will increase and may create a fake transient event. It is therefore
necessary to correct the time intervals for these acceptance variations.

\subsection{Acceptance estimation}

\begin{center}
\begin{figure}[t]
\centering
    \subfigure{\includegraphics[width=9cm]{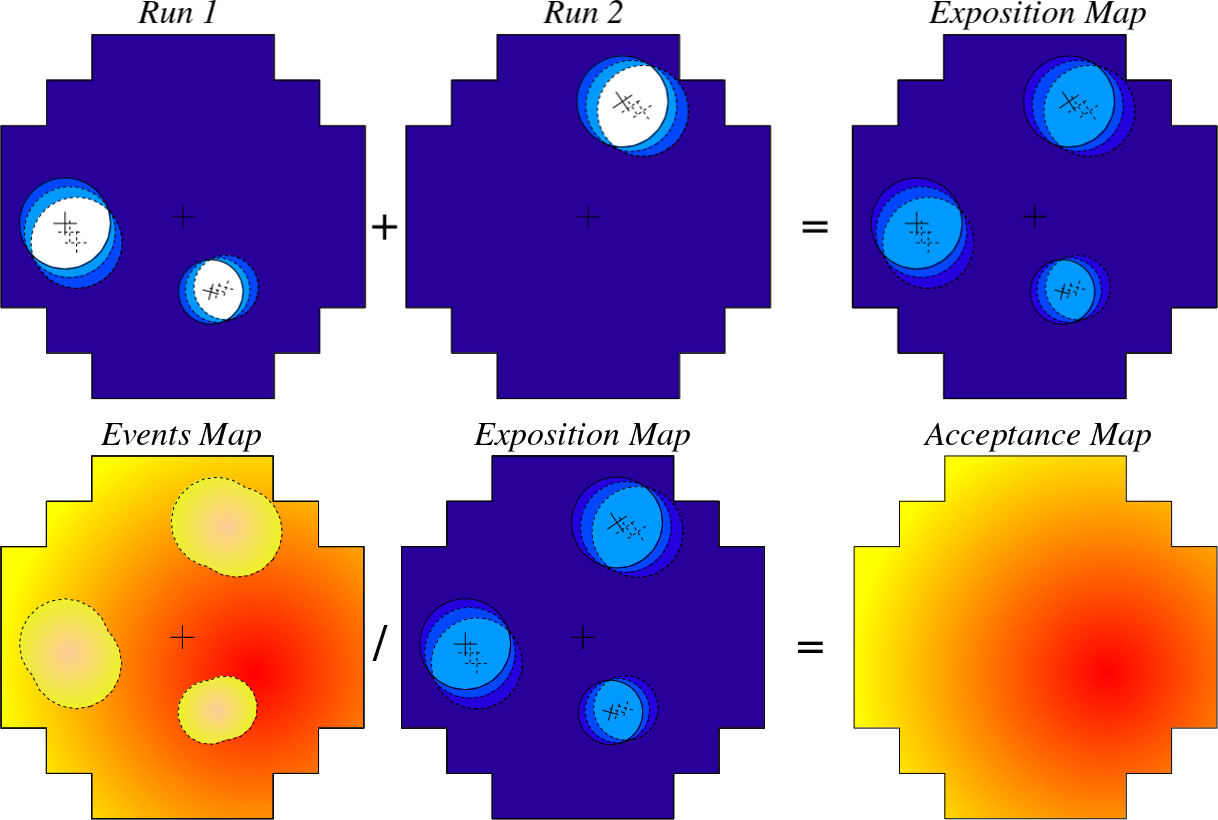}}
\caption[]{illustration of the procedure to compute the \emph{acceptance map} in the
  detector's frame with two observation runs and two exclusion regions in the
  first run and one in the second run. \emph{Top :} the \emph{exposition map} in
  the detector frame, which represents the fraction of time each position is
  outside any exclusion region, is computed by summing the \emph{exposition maps} of
  each individual observation run (weighted by the total number of events
  outside exclusion regions). \emph{Bottom :} the \emph{event map} is filled
  for the two runs, by the gamma-like events outside \emph{exclusion regions}.
  Dividing the \emph{event map} by the \emph{exposition map}, gives the
  \emph{acceptance map} which -- once reprojected on the sky and properly normalized -- provides the
  expected number of gamma-like background events at each position on the sky.}
\label{fig:acceptancemap}
\end{figure}
\end{center}

In order to correct for the effect of acceptance variations, the instrument response is determined for
each (observed) position on the sky and for each time during the observations in
the following way.  First, an \textit{exposition map} is calculated in the
detector frame. This map represents the fraction of time each position is
outside regions known to contain gamma-ray sources, which are not used for
background estimation and are hereafter called \textit{exclusion regions}.  As
the sky is moving in the detector frame, the trajectory of each
\textit{exclusion region} in this frame has to be taken into account for a
proper determination of the exposure. This can be done by using the fraction of
hadron-like events in exclusion regions as a function of the position in the
detector frame. Then, the \textit{exposition maps} of all observation runs are
summed up, weighted by the total number of (non-excluded) events in each run. At
the same time, the map of gamma-like events outside any exclusion regions in the
detector's frame are summed up. Dividing this map of events by the
\textit{exposition map} gives an \textit{acceptance map} in the detector's
frame, as illustrated in Fig. \ref{fig:acceptancemap}. This map can be
re-projected on the sky to determine, for the given set of observation runs, the
expected rate of gamma-like background events at a given position in the
sky. Zenith angle effects are taken into account by applying this procedure in
predefined zenith angle bands. Finally, once the acceptance map in the
detector's frame is computed, the acceptance at a given sky position and
observation time can be determined. It is done by retrieving the acceptance
value (in the detector's frame) at the position corresponding to that of the
requested sky position and time, and by scaling it by the gamma-like event rate
of the corresponding observation run.  The instrument response in
this case is then the expected instantaneous event rate from gamma-like
events. It is used in the following parts to correct the observed time intervals for
the acceptance variations.

Even though it depends on the observation setup, observation conditions
and analysis performances, typically a few ($\sim 3 - 5$) hours of observation is
needed to be able to correctly estimate the acceptance from the data. In
particular, it is technically necessary to observe a test position with
different wobble offset to avoid the exclusion regions to always stand at the
same position in the detector's frame. In case this method can not be used (for
instance for shorter observation times such as a single observation run), it is
possible to assume a radially symmetric acceptance and estimate it using tables
(also binned in zenith angle) built from dedicated observations. This should
lead to a slightly less accurate estimate of the acceptance but for both cases,
the acceptance can be determined at the level of a few percent which is
sufficient when dealing with the level of fluxes that can be probed (see Sect.
\ref{section:fakeinjection}).

\subsection{Acceptance correction}

Once the acceptance has been estimated, the time intervals can be corrected for
the instrumental induced variations. The gaps between observation runs are
taken into account by only considering the time intervals between detected
events in the same observation run.  The acceptance-corrected time intervals $\mathrm{d}\tau$ can be computed from the observed time intervals $\mathrm{d}t$ and the acceptance $a(t)$ with the formula : $\mathrm{d}\tau = a(t) \mathrm{d}t$.
For a given set of $N$ recorded times $\lbrace t_i\rbrace_{i=1\ldots N}$, the
corrected time intervals can be expressed as follows (taking the approximated acceptance value at the middle of the time interval):

\beq
\Delta \tau_i = \tau_{i+1} - \tau_i = \frac{a(t_i)+a(t_{i+1})}{2} \Delta t_i
\eeq

where $\Delta t_i = t_{i+i}-t_i$.

The acceptance-corrected time intervals can then be computed using the
integrated acceptance. If we note $\tilde{a}_i$ the integral of the acceptance
from the beginning of the observation to the time $t_i$ $\left(\tilde{a}_i =
\int_{t_0}^{t_i} a(t) \mathrm{d}t\right)$, then :

\begin{eqnarray*}
  \tilde{a}_{i+1} & = & \int_{t_0}^{t_{i+1}} a(t) \mathrm{d}t \\
  & = & \int_{t_0}^{t_{i}} a(t) \mathrm{d}t + \int_{t_i}^{t_{i+1}} a(t) \mathrm{d}t \\
  & \simeq & \tilde{a}_{i} + \frac{a(t_i)+a(t_{i+1})}{2}\Delta t_i \\
  & \simeq & \tilde{a}_{i} + \Delta \tau_i
\end{eqnarray*}

then : $ \Delta \tau_i = \tilde{a}_{i+1} - \tilde{a}_{i} $.

The obtained set of acceptance-corrected time intervals $\lbrace\Delta
\tau_i\rbrace_{i=1\ldots N-1}$ can be normalized to have a mean value of $1$ in
the following way : if $N$ is the total number of events, then $ \int a(t) \mathrm{d}t =
\int \mathrm{d}\tau = \beta N $, which, for discrete times is : $ \sum \Delta \tau =
\beta N $. The normalization factor $\beta $ is thus calculated by comparing the
sum of the corrected time intervals and the number of events. The normalized
acceptance-corrected time intervals are then : $\hat{\Delta \tau} = 1/\beta
\cdot \Delta \tau$.  The expected corrected time intervals distribution is then
an exponential with an expected slope of $1$. This has proven to be helpful to
check whether the procedure is working correctly.

Once this step is performed, the normalized acceptance-corrected time series can
be constructed from the sum of the normalized acceptance-corrected time
intervals. The time span between two observations is taken care of by
suppressing the gap between the last event of an observation and the first event
of the next one. The total time series is therefore built from the time
intervals between consecutive events, purposely removing the intervals between
events from different observations. The acceptance being estimated and corrected
for each observation, the presence of gaps and their size has no influence on
the final time series. In the case of a steady source or background, the time
series constructed following this procedure follows a Poisson process with a
steady event rate of 1, no matter the observation conditions or observation setup.
Statistical tests can be applied to this time series to search for
transient events. In addition, the acceptance determination being possible on
the whole field-of-view, this allows to perform blind searches for transient
events at any observed position on the sky.

\subsection{Event selection and Map generation}

The arrival date of the recorded events are stored during the analysis together
with the information on the direction in the sky. Given these information and
applying the acceptance correction procedure described above, a list of arrival
times is built for a direction in the sky within a given integration
radius. With IACT, observations are usually separated in observation runs of
typically $30$ minutes. For a given source or a given region in the sky,
consecutive observation runs can be separated by a time ranging from a few
minutes to a few years - emphasizing the need for an acceptance estimation and
correction procedure. By construction, the set of acceptance-corrected arrival times has the
following properties : the mean time interval between two events is equal to $1$
and the arrival time of the last event is equal to the number of events in the
set.

For map generation, a set of acceptance-corrected arrival times is determined at each sky
position of a pre-defined grid (defining pixels in the map). In the
applications described below, we applied an integration radius
which is the same as the $\theta^2$ cut optimizing the detection of point-like
sources (see~\citet{rollanddenaurois} for more details on this aspect). In order
to avoid statistical bias and as mentioned in the next section, we
also apply the condition that at least 20 events are in the sample at each sky
position.

\section{Variability tests description} \label{section:tests}

Several tests for small time scales flux variations are available in the
litterature.  In this article, we describe and use three tests, original or
adapted from the litterature, which apply to the acceptance-corrected event times
if not specified otherwise. As the tests could also apply to the raw measured
times in case of a constant acceptance, they are described in the
following sections using the generic notation $T$ for the times.

\subsection{An adaptation of the Exp-Test}

The Running Exp-Test is built upon the Exp-Test described in
\citet{exptest}. For the convenience of the reader, this test is reminded here
and the Running Exp-Test is then described.

\subsubsection*{Exp-Test}

For event arrival times $\xi_i$ following Poisson statistics, there is a constant
$C$ such that for any $\Delta \xi > 0$ dividing the total time interval $\Xi$ in
evenly spaced intervals $\Delta \xi$, the number of events per interval follow a
Poisson law with $\lambda = \Delta \xi /C$ (the Poisson distribution is
$P_\lambda(n) = e^{-\lambda}\cdot\frac{\lambda^n}{n!}$, where $\lambda$ is the
expectation value).  The probability density function of time intervals $\Delta
\xi$ is then a decreasing exponential :

\beq
f_C(\Delta \xi) = \frac{1}{C}\cdot \exp\left( -\frac{\Delta \xi}{C}\right)
\eeq

For events observed at times $(T_i)_{i = 1\ldots N+1}$, the distribution of time
intervals between two consecutive events is :

\beq
\lbrace\Delta T_i\rbrace _{i=1\ldots N} := \lbrace (T_{i+1}-T_i)\rbrace_{i=1\ldots N}
\eeq

with a mean value $\overline{\Delta T} := C^\ast$. If the constant $C$ is fixed
at the observed value, the $\Delta T_i$ then follow the distribution
$f_{C^\ast}(\Delta \xi)$. A test can then be applied on the observed $\Delta T_i$
to determine whether they follow the distribution $f_{C^\ast}(\Delta \xi)$ The
estimator built by \citet{exptest} is the following :

\beq
M(F) := \int_0^{C^\ast}\left(1-\frac{\Delta \xi}{C^\ast}\right)\cdot F(\Delta \xi)~d \Delta \xi
\eeq

where $C^\ast = \int \Delta \xi \,F(\Delta \xi)\,d\Delta \xi$ and $F(\Delta \xi)$, which
correspond to the fraction of intervals equals to $\Delta \xi$ is defined as :

\beq
F(\Delta \xi):= \frac{1}{N}\sum_{i=1}^N \delta(\Delta \xi - \Delta T_i)
\eeq

The estimator can be expressed in the following way :

\beq
M = \frac{1}{N} \sum_{\Delta T_i < C^\ast} \left(1 - \frac{\Delta T_i}{C^\ast}\right)
\eeq

It has the property to be equal to $1/e$ for $F = f_{C^\ast}$, which is the
expected value for a Poisson behaviour. The estimator will be greater than the
expected value for a burst-like behaviour (excess of small $\Delta T_i$) and
smaller than the expected value for a periodic behaviour.

This estimator can be normalized to follow a normal distribution :

\beq
M_r = \frac{M - (1/e - \alpha / N)}{\beta / \sqrt{N}}
\eeq

where $N$ is the number of events and the values of $\alpha = 0.189 \pm 0.004$
and $\beta = 0.2427 \pm 0.0002$ were adjusted in \citet{exptest} by the
means of simulations, assuming a steady Poisson process. In the same
publication, it is noted that at least 20 events should be in the sample in
order to avoid statistical bias. It is therefore possible to express the result
of the test in terms of number of standard deviations or significance.

\subsubsection*{Running Exp-Test}

For a transient event of fixed duration, it is expected that the detection
efficiency of the Exp-Test is reduced if the total observation time
increases. The fixed number of "smaller-than-expected" time intervals
corresponding to the flare may be significant in a small dataset while, as the
dataset increases, it would be hidden in the global distribution of time
intervals expected from the steady Poisson process.

The Running Exp-Test aims at resolving this issue by applying the same estimator
as for the Exp-Test to subsets of the total number of events. In practice, the
number of events on which the test is performed is fixed to a value chosen by
the user which can be physically motivated (the number of events corresponds to
a time scale that may be expected from a given physical process). The Exp-Test
is performed on a sliding window of the chosen size (in number of events), the
starting event of each test window being increased by one at each step. For each
window, the test is performed with the mean value of the time interval
($C^\ast$) of the whole dataset.  The result of the Running Exp-Test is then a
set of significances and we choose to use the maximum value of this set as the
output of the test.

As the test windows overlap, they are not independent and trial factors have to
be corrected for. This is done with simulations : for a given dataset, the same
number of events is simulated assuming a pure Poisson behaviour and the Running
Exp-Test is applied. The simulation is performed a large number of times, giving
a distribution of expected values for a pure Poisson process. From this
distribution, the significance of the Running Exp-Test obtained on the real
dataset can be corrected (using the cumulative distribution for instance).

The advantages of this test is that its efficiency is much less affected by the
size of the dataset and can be adapted to the physical process one aims to
detect. On the other hand, a window size has to be chosen by the user which will
have an impact on the results of the test. Due to the acceptance variations and the
Poisson process itself, this window size does not correspond to a constant
time-scale probed.

\subsection{ON-OFF Test}

The ON-OFF test for time series was developed in order to probe variability on a
given time scale without it being affected by the acceptance variations. With this
test, the probed time scale is the real physical one and it is not affected by
the acceptance correction procedure. This
method -- similar to the method used at high energies by the Fermi-LAT
\citep{FAVA} -- is the analogous in the time domain to the standard ON-OFF
method used to compute excess and significance maps~\citep{bergebg} in very high
energy gamma-ray astronomy.

\begin{center}
\begin{figure}[htb!]
\centering
\includegraphics[width=7cm]{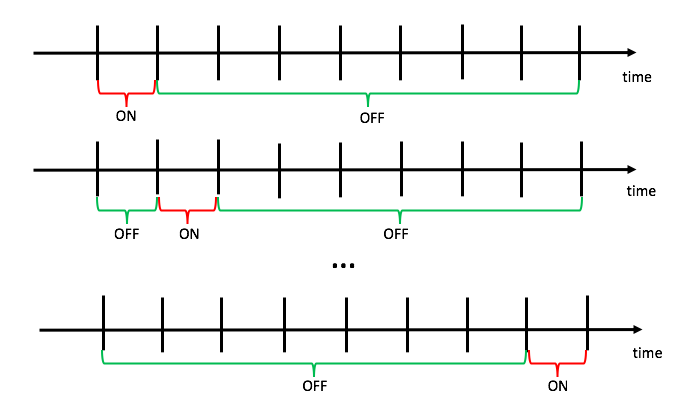}
\caption[Scheme of ON-OFF test. The first line considers the first subinterval
  as ON region and the others as OFF regions. The second line does the same for
  the second subinterval. The last line considers the last bin as ON region. See
  text for more details.]{Scheme of ON-OFF test. The first line considers the
  first subinterval as ON region and the others as OFF regions. The second line
  does the same for the second subinterval. The last line considers the last bin
  as ON region. See text for more details.}
\label{ON-OFFscheme}
\end{figure}
\end{center}

The set of events arrival times is binned with a binning size being the probed
time scale (the first event defines the starting time of the first bin). The
times considered here are the real ones and not the acceptance-corrected ones.
The number of events in each time bin is compared to the number of
events in all the others bins as shown in Figure \ref{ON-OFFscheme}. The number
of events in a bin is composed of background events, possibly gamma-rays if we
are looking at a steady source, and additional gamma-rays if an increase in the
source activity occurs or if a transient source appears. The "steady" number of
events in a considered bin is estimated using all the other bins. The number of
gamma-rays, also called excess of gamma-rays, is therefore computed in a given
bin (the ON region) with respect to all the other bins (the OFF region). The
significance of this excess can then be computed from eq. 17 of the Li\&Ma
publication \citep{lima}. The OFF region should not contain bins in which a
significant excess is measured. The procedure is thus iterative and bins
containing significant signal (above 5\,$\sigma$) are not used for background
estimation.
As prescribed by \citet{lima}, there should be at least 10 events in the
ON and OFF regions. In order to declare a detection significant, one can follow
the rather conservative prescription from \citet{CTAMC} : the significance from
eq. 17 of \citet{lima}, the number of excess events and the number of excess
over background events should be above 5, 10 and 0.05, respectively. The latter
value being mainly guided by the level of control of the systematics. The
corresponding minimal timescale that can be probed depends on the instrument and
the overall performances of the analysis. In the application described in this
publication, the minimal timescale is $\sim 2$ minutes.

For N time bins, the excess of gamma-rays in a given time bin is $N_{\gamma,i} =
N_{ON,i} - \alpha_{i} \times N_{OFF,i}$, where :

\begin{itemize}
    \item $N_{ON,i}$ is the number of events in the i$^{th}$ bin,
    \item $N_{OFF,i}$ is the total number of events outside of the i$^{th}$ bin
      (and from bins which have not been recognized as containing signal),
    \item $\alpha_{i}$ is the ratio of the total acceptance in the ON and in the
      OFF bins.
\end{itemize}

\begin{center}
\begin{figure*}[t]
\centering
    \subfigure{\includegraphics[width=0.8\textwidth]{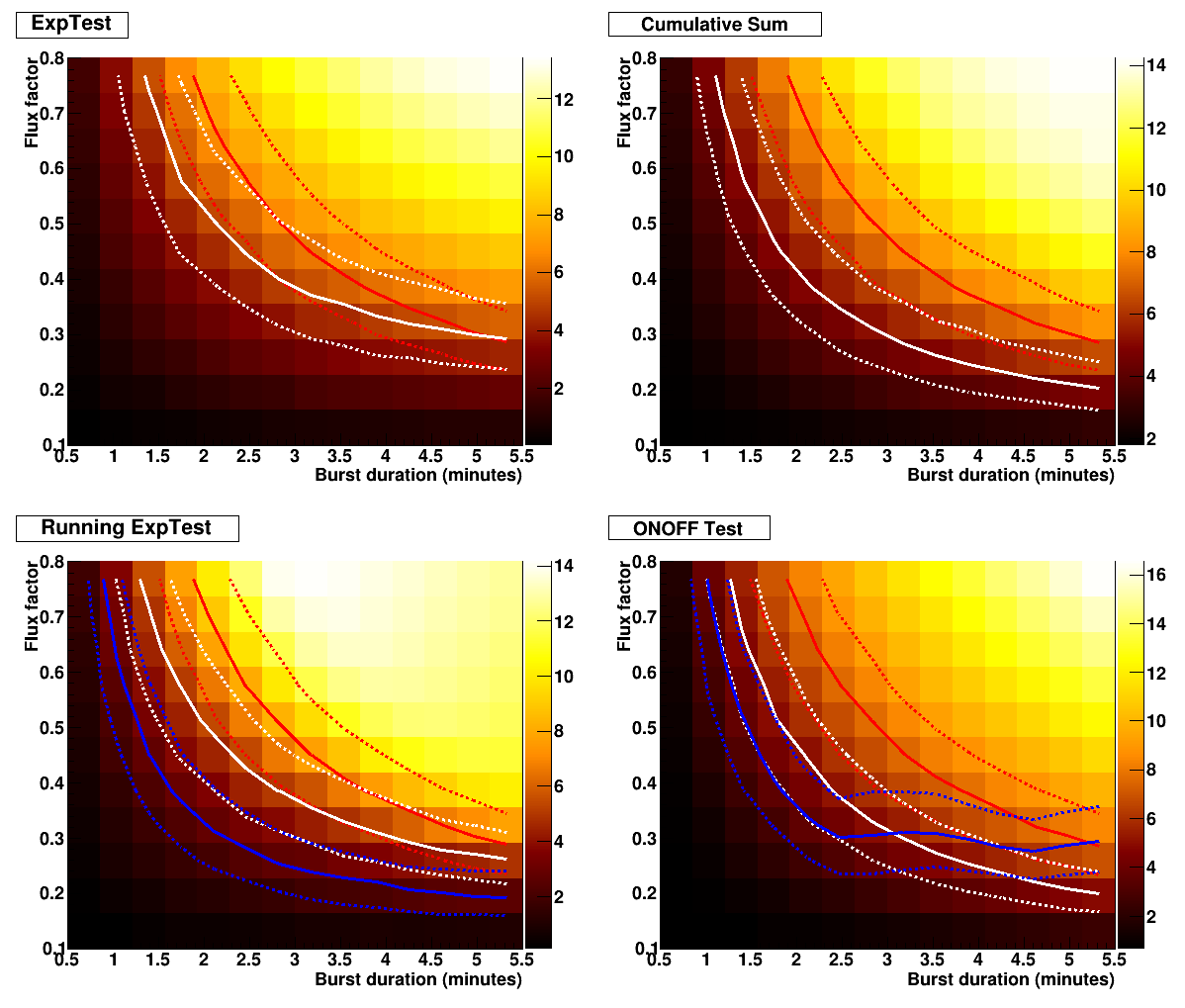}}
\caption[]{Exp-Test, Cumulative Sum, Running Exp-Test and ON-OFF Test
  performances for a run duration of 28 minutes and an off-axis angle of
  $0.5^\circ$. Simulations are performed as described in the text for burst
  durations between 0.5 and 5.5 minutes and flux scaling factor between 0.1 and
  0.8.  The map represents the significance of the tests, the white lines
  corresponds to the position of the $5\sigma$ boundary (the dotted line
  corresponding to the $1\sigma$ contours) and the red lines corresponds to the
  $5\sigma$ significance boundary of the standard Li\&Ma test. For the Running
  Exp-Test, the number of events in the window is set to 20 (map and blue lines)
  and 50 (white lines). For the ON-OFF Test, the probed time-scales are 2
  minutes (map and blue lines) and 5 minutes (white lines).}
\label{fig:simu_test}
\end{figure*}
\end{center}

Again, we consider the maximum significance value over all the bins as the
output of the test. Since a number of bins is tested for an excess, trial
factors should be taken into account. The final, post-trials
significance value is the pre-trials significance of the excess (determined
after the iterative procedure) and corrected by the number of time bins tested.
Since the significance follows a normal distribution in the absence of signal,
the pre-trials significance can be converted into a probability, corrected for
trial factors (using $P_{post} = 1 - (1- P_{pre})^{N_{trials}}$) and converted
back in a post-trials significance value.

An additional step can be achieved with this test : knowing the observation
conditions in each time bin, the number of expected events can be computed from
Monte-Carlo simulations. Under the assumption of a spectral shape for a putative
gamma-ray emitter, this number can be compared to the observed excess in this
bin and a value (or an upper limit) for the flux contained in this bin can be
derived. This value (or upper-limit) can be interpreted as the additional flux
due to the transient process for a steady source or as the flux of a source
emitting in the considered energy range only during the outburst.

Even though a time-scale has to be chosen by the user, this test has the
advantage of allowing for a real time-scale to be probed (as opposed to an
acceptance corrected time scale for the Running Exp-Test). It relies on a robust
method and can easily be used to the estimation of physical parameters (flux or
upper-limit on the flux).

\subsection{Cumulative Sum test}

The Cumulative Sum test is originally a method to detect and monitor change points, often
used in the industry for quality control. It was first developped in \cite{CuSumChart}.
The version proposed here is based on the fact that the statistics of the time
interval between two events is known for a Poisson behaviour. The principle is
to compute the cumulative sum of the time intervals, retrieving at each step the
global mean value in order to get a variable with a null mean value.

For a set of time intervals, ordered in time $ \lbrace\Delta T_i\rbrace
_{i=1\ldots N} := \lbrace(T_{i+1}-T_i)\rbrace_{i=1\ldots N} $, one can compute
the cumulative sum variable, which is for the time i :
\beq
\chi_i = \sum_{k = 1}^i \left( \Delta T_k - \langle \Delta T \rangle
\right)
\eeq

where $\langle \Delta T \rangle = 1/N \sum_{i=1}^N \Delta T_i = C$.  This estimator has a null mean as for each i :

 \begin{eqnarray*}
 \langle \chi_i \rangle &=& \sum_{k=1}^i \left( \langle \Delta T_k
    \rangle - \frac{1}{N}\sum_{j=1}^N \langle \Delta T_j \rangle
  \right) \\
  &=& 0
 \end{eqnarray*}

If the events follow a Poisson behaviour, then :

\beq
\langle \Delta T^2 \rangle = \int \Delta T^2 f_C(\Delta T)\, \mathrm{d}\Delta T = 2 C^2
\eeq

The variance of the $\chi_i $ variable is then :

\begin{eqnarray*}
Var(\chi_i) &=& \langle \chi_i^2 \rangle - \langle \chi_i \rangle^2 = \langle \chi_i^2 \rangle \\
 &=& \bigg\langle \left( \sum_{k=1}^i \Delta T_k - \frac{i}{N} \sum_{j=1}^N \Delta T_j \right)^2 \bigg\rangle \\
 &=& \frac{iC^2}{N}(N-i) \\
 \end{eqnarray*}

The complete calculation can be found in the Appendix A.  It is thus possible to
determine at each step if the variable is close to the distribution expected for
a steady source and compute the number of standard deviations from it.
Since the estimator is bound to be null for the first and the last event,
a reasonable minimum number of events should be used to avoid statistical bias.
As such, the prescription from \citet{exptest} for the Exp-Test to have at least
20 events in the sample appears to be a safe choice in this case as well.

The advantage of this test is its fast computing time and the fact that it does
not rely on any assumption regarding the time-scale to probe.

%____________________________________________________________________________________

\section{Performance of the tests}
\label{section:fakeinjection}

In order to test and compare the performances of the methods described in the
section \ref{section:tests}, a simple simulation is adopted. It is based on the
knowledge of two quantities for a given observation run : the background rate
and the expected rate of gamma rays coming from a source with a given spectrum.

\begin{center}
\begin{figure*}[htb!]
\centering
    \subfigure{\includegraphics[width=0.4\textwidth]{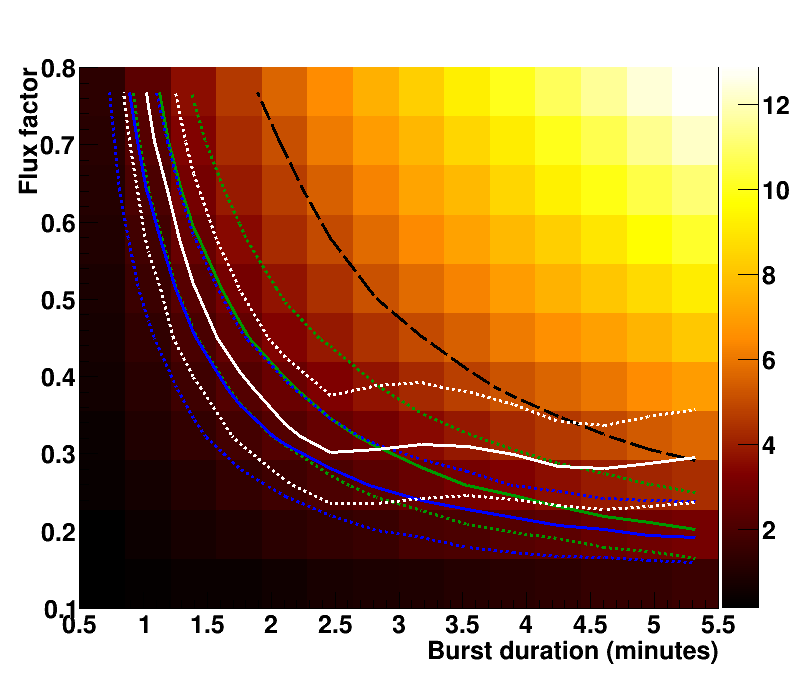}}
    \subfigure{\includegraphics[width=0.4\textwidth]{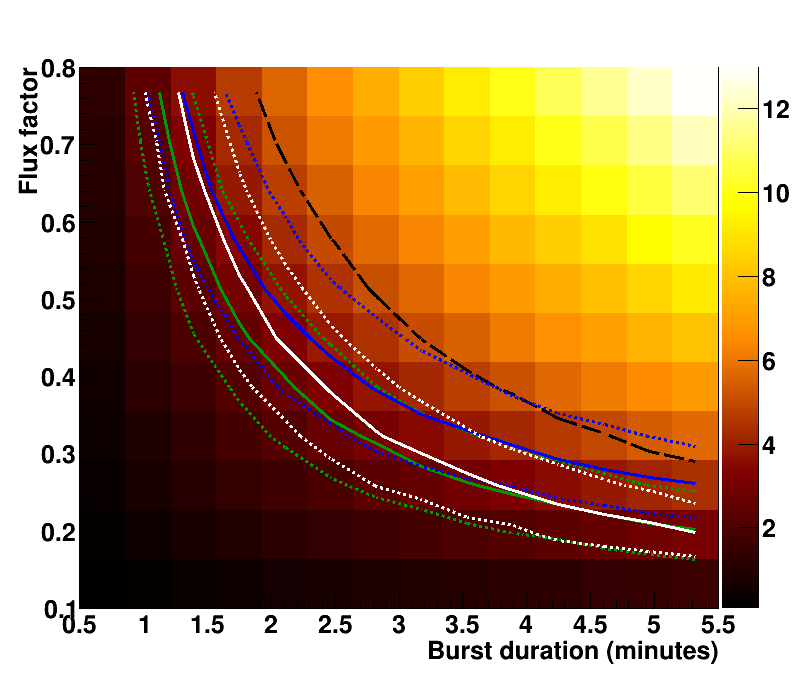}}
\caption[]{Comparison of the Cumulative Sum Test, the Running Exp-Test and the
  ON-OFF Test for a run duration of 28 minutes and an off-axis angle of
  $0.5^\circ$. Similarly to Fig.\ref{fig:simu_test}, simulations are performed
  for burst durations between 0.5 and 5.5 minutes and flux scaling factor
  between 0.1 and 0.8. The color maps represents the Li\&Ma significance with
  the $5\sigma$ boundary line in black dashed. The $5\sigma$ boundary lines (with their
  confidence contours) are represented in white for the ON-OFF Test (with probed
  time-scale of 2 minutes (left) and 5 minutes (right)), in green for the
  Cumulative sum test and in blue for the Running Exp-Test (with a running
  window of 20 events (left) and 50 events (right)). }
\label{fig:simu_test_comparison}
\end{figure*}
\end{center}

\subsection{Simulation procedure}

From a real observation run, obtained with H.E.S.S. II in combined mode
\citep[i.e. using either monoscopic events from CT5 only or events seen with at
  least two telescopes,][]{Combined} with the \emph{Loose} set of cuts, we
derived the average rate of gamma-like background events (see Sect.
 \ref{section:acceptance}) as a function of the angular distance to the
 observation position (the off-axis angle). The derived rate depends on
the observation conditions and the telescope performances. In our case, the
observation run was taken towards an empty extragalactic field with a mean
zenith angle of $\sim 10^\circ$.
We also assume that the acceptance is constant during the run.  This is the case
when applying the method described in section \ref{section:acceptance} and this
is in any case a valid assumption if the observation is not too long and if the
observation conditions are stable, which is supposed to be the case here.
If a systematic shift in the determination of the acceptance occurs,
this would be equivalent to a change in the event rate over the whole
observation. By construction this would have no effect for the detection power
of the Exp-Test, the Running Exp-Test or the Cumulative Sum test. Only the
ON-OFF Test would have its sensitivity modified at the same level as the
amplitude of the shift (which is not expected to happen at a level of more than
a few percent). If the acceptance is incorrectly estimated and if this error in
addition varies with time (for instance due to zenith angle gradient not
properly taken into account), all the tests could be affected. However, this
would still be at the level of a few percent and would only have a moderate
effect on their detection efficiency.

From the instrument response function, and for a given assumed spectrum, we can
also derive the rate of events expected from a source of gamma-rays at a given
off-axis angle in this observation run. In the following, we use a source with a
power-law spectrum ($\mathrm{d}N/\mathrm{d}E = \phi(E_0) \times (E / E_0)^\Gamma$) with the
parameters of the Crab nebula as measured in \citet{Crab2006} ($\phi
(1~\mathrm{TeV}) = 3.45 \times 10^{-11} \mathrm{cm^{-2}s^{-1}TeV^{-1}}$ and
$\Gamma = -2.63$). Since the event rate scales linearly with the flux, we
compute this rate once and only scale the value in order to test for different
normalizations.

With the knowledge of the background rate and the gamma rate (of a source with
an assumed spectrum), we perform the simulations in the following way :
\begin{itemize}
    \item We chose the duration of the observation run to 28 minutes (standard
      duration in H.E.S.S.) and an off-axis angle value. From the average
      background rate determined above, and under the assumption of a Poisson
      behaviour, we generate random time intervals between consecutive events and build the
      time series of gamma-like events.
    \item In addition to the gamma-like background, we assume that a burst of
      gamma rays from a transient source is observed, the lightcurve of which is
      a step-function\footnote{This hypothesis is well suited to study
      and compare the performances of the tests in a simple case. For a more
      complex emission model like the afterglow phase of a GRB or a
      generalized-gaussian shaped peak to study the sensitivity to variability
      in AGNs, dedicated studies should be performed.}. The source is assumed
      to be active during a given amount of time and
      quiet during the rest of the run. We chose the duration and the flux level
      of the burst and, as for the gamma like events, we determine the time
      series of gamma ray events from the transient source using the expected
      rate of gamma-rays.
    \item The two time series are then merged and the tests are applied to the
      global dataset.
\end{itemize}

Apart from observation run and analysis configuration, the parameters for
the simulations are the off-axis angle, the run duration, the duration of the
burst and its level of flux. For the tests, only the Running Exp-Test and the
ON-OFF Test have a parameter that needs to be chosen (the number of events in
the running window and the time-scale to probe for the Running Exp-Test and the
ON-OFF Test respectively).

When simulating the gamma-like events, we simultaneously perform 10 different realisations of
the simulation\footnote{Since we draw the time intervals from the average rate
  assuming a Poisson process, each simulation provides a different number of events.}.
These additional simulations serve as a background estimate to
derive the significance of the source as would be done for a standard analysis
using the reflected background method \citep{bergebg}.

\subsection{Results}

In order to perform the tests and compare them, we set the run duration to 28
minutes and the off-axis angle to $0.5^\circ$. We then perform simulations as
described above for different pairs of burst durations and flux scaling factor
values. The tested values range from 0.5 to 5.5 minutes for the burst duration
and from 0.1 to 0.8 for the flux scaling factor. The number of events in the
running window for the Running Exp-Test is set to 20 and 50 and the time-scales
probed by the ON-OFF Test are 2 and 5 minutes.

For each pair of values, we perform 1000 different realisations of the simulation. We obtain, for each test,
a distribution of 1000 significance values, of which we derive the mean and the
RMS. It is expected that at short (long) burst durations and for low (high) flux
scaling factors, the significance of the test is low (high). The results
obtained from the simulation presented here are compatible with those obtained
using the \textit{Injector} module which was developed and implemented in the
H.E.S.S. analysis software to test the transient analysis tools performances
with any kind of spectra and lightcurves ~\citep[][]{fakeeventsinjector}.

In figure \ref{fig:simu_test}, the mean significance is represented as a
function of the burst duration and the flux scaling factor. The white lines in
figure \ref{fig:simu_test} represent the $5\sigma$ contour and the white dashed
lines represent $5\sigma$ contours when taking into account the RMS of the significance distribution. Similarly,
the red lines show the $5\sigma$ contours of the significance obtained with
eq. 17 of the Li\&Ma publication \citep{lima}. On the bottom maps of figure
\ref{fig:simu_test}, the blue lines represent the $5\sigma$ contour of the
Running Exp-Test (left) and the ON-OFF Test for different values of probed
time-scales (see caption for details).

In figure \ref{fig:simu_test_comparison}, the $5\sigma$ contours of the
Cumulative Sum Test, the Running Exp-Test and the ON-OFF Test are represented on
the same figure. The Running Exp-Test appears to be the most sensitive test but,
like the ON-OFF Test, its result depends on an a-priori choice of time scale to
be probed. The Cumulative Sum Test gives results similar to the other two tests
but without any assumption on the time-scale. In any case, all of the tests
presented in this publication perform significantly better than the standard
method used for source detection for time scales shorter than approximately $8$
minutes.

\section{Conclusion}

New methods were developed to search for transient events in VHE
gamma-ray data from IACTs. The methods described in this paper allow to estimate
the instrument response at any time and position of the observations. This in
turns allows to apply statistical tests on the acceptance-corrected time
series. The tests can be applied at any observed position, leading to the
possibility to perform blind search of transient sources over the whole observed
field-of-view.

Several tests were presented and, when applied to short time scale signals, they
all perform significantly better than the standard Li\&Ma test. The better sensitivity
to transient events achieved with these methods show a clear interest to use
them for weak transient sources searches. In addition, these methods allow to
search for transient emissions such as prompt emission from GRBs in archival
data which could not be revealed by standard analysis methods.

The statistical tests described in this publication are already used for the
analysis of H.E.S.S. data and will be of prime interest for CTA. They will be
implemented in the analysis softwares of CTA :
ctools \citep{ctools} and
gammapy \citep{Gammapy}. The performance
of the methods described in this publication applied to CTA analysis and for
more realistic cases will be estimated in a forthcoming publication.

\section*{Acknowledgements}

We thank Dr. M. de Naurois, spokesman of the H.E.S.S. Collaboration and Prof.
O. Reimer, chairman of the H.E.S.S. Collaboration board, for allowing us to use
H.E.S.S. data in this publication.
F.B. thanks M. Lemoine-Goumard for supporting this work and for very fruitful
discussions.
We are also grateful to V. Marandon, D. Sanchez, A. Fiasson and J.-F.
Glicenstein for their careful reading of the manuscript and for providing us
with useful suggestions.
Finally, we would like to thank all the members of the H.E.S.S. Collaboration
for their technical support and for helpful discussions.

\def\aap{A\&A}
\def\apj{ApJ}%
\def\apjl{ApJ}%
\def\apjs{ApJS}%
\def\nar{New A Rev.}%
\def\nat{Nature}%
\bibliography{main}

%__________________________________________________________________

\appendix
\section{Cumulative Sum variance derivation}

The variance of the Cumulative Sum estimator described in section \ref{section:tests} is given by :

\begin{align*}
  Var(\chi_i) &= \langle \chi_i^2 \rangle - \langle \chi_i \rangle^2 = \langle \chi_i^2 \rangle \\
  &= \bigg\langle \left( \sum_{k=1}^i \Delta T_k - \frac{i}{N} \sum_{j=1}^N \Delta T_j
  \right)^2 \bigg\rangle \\
  &=
  \begin{aligned}[t]
  &\bigg\langle \sum_{k=1}^i \Delta T_k \sum_{j=1}^i \Delta T_j - \frac{2i}{N} \left( \sum_{k=1}^i \Delta T_k \sum_{j=1}^N \Delta T_j \right) \\
  &+  \frac{i^2}{N^2} \left( \sum_{k=1}^N \Delta T_k \sum_{j=1}^N \Delta T_j \right) \bigg\rangle \\
  \end{aligned}
  \end{align*}

which gives :
\begin{align*}
  Var(\chi_i) &=
  \begin{aligned}[t]
  &\bigg\langle \sum_{k=1}^i \Delta T_k \sum_{j=1}^i \Delta T_j \bigg\rangle - \frac{2i}{N} \bigg\langle \left( \sum_{k=1}^i \Delta T_k \sum_{j=1}^N \Delta T_j \right) \bigg\rangle \\
  &+  \frac{i^2}{N^2} \bigg\langle \left( \sum_{k=1}^N \Delta T_k \sum_{j=1}^N \Delta T_j \right) \bigg\rangle\\
  \end{aligned}
\end{align*}

The three terms in this sum are :

\begin{eqnarray*}
 \bigg\langle \sum_{k=1}^i \Delta T_k \sum_{j=1}^i \Delta T_j \bigg\rangle &=& \sum_{k=1}^i \sum_{j=1}^i \langle \Delta T_k \Delta T_j \rangle \\
  &=& \sum_{k=1}^i \sum_{j=1, j\ne k}^i \langle \Delta T_k \rangle \langle \Delta T_j \rangle + \sum_{k=1}^i \langle \Delta T_k^2 \rangle\\
  &=& i(i-1)C^2 + 2iC^2\\
\end{eqnarray*}

\begin{eqnarray*}
  \bigg\langle \sum_{k=1}^N \Delta T_k \sum_{j=1}^N \Delta T_j \bigg\rangle &=& N(N-1)C^2 + 2NC^2\\
\end{eqnarray*}

\begin{eqnarray*}
  \bigg\langle \sum_{k=1}^i \Delta T_k \sum_{j=1}^N \Delta T_j \bigg\rangle &=& \sum_{k=1}^i \sum_{j=1}^i \langle \Delta T_k \Delta T_j \rangle + \sum_{k=1}^i \sum_{j=i+1}^i \langle \Delta T_k \Delta T_j \rangle\\
  &=& i(i-1)C^2 + 2iC^2 + i(N-i)C^2 \\
\end{eqnarray*}

Putting everything together gives :

\begin{eqnarray*}
  Var(\chi_i) &=& \frac{iC^2}{N}(N-i) \\
\end{eqnarray*}

\end{document}